\documentclass[prl,onecolumn,showpacs,preprintnumbers,amsmath,amssymb,floatfix]{revtex4}

\usepackage[dvips]{graphics,graphicx,color}

\begin{document}

\title{Phase Diagram of the Low-Density Two-Dimensional Homogeneous Electron
Gas}

\author{N.\ D.\ Drummond and R.\ J.\ Needs}

\affiliation{TCM Group, Cavendish Laboratory, University of Cambridge, J.\ J.\
Thomson Avenue, Cambridge CB3 0HE, United Kingdom}

\begin{abstract}
We have used quantum Monte Carlo methods to calculate the
zero-temperature phase diagram of the two-dimensional homogeneous
electron gas.  We find a transition from a paramagnetic fluid to an
antiferromagnetic triangular Wigner crystal at density parameter
$r_s=31(1)$ a.u.\ and a transition to a ferromagnetic crystal at
$r_s=38(5)$ a.u.  The fully spin-polarized fluid is never stable.  We
searched for, but did not find, the ferromagnetic ``hybrid'' phase
proposed by H.\ Falakshahi and X.\ Waintal [Phys.\ Rev.\ Lett.\
\textbf{94}, 046801 (2005)].
\end{abstract}


\maketitle

The two-dimensional (2D) homogeneous electron gas (HEG) is one of the most
important model systems in condensed-matter physics.  It is our basic model of
the conduction electrons in layered semiconductor devices, and consists of a
set of electrons moving in 2D in a uniform, inert, neutralizing background.
At high densities the HEG exists in a Fermi fluid phase, but at low densities
it forms a Wigner crystal to minimize the electrostatic repulsion between the
electrons \cite{wigner}.  Classical 2D Wigner crystals can be produced in the
laboratory by spraying electrons onto droplets of liquid He \cite{grimes}, and
quantum Wigner crystals can be formed at the interface between two
semiconductors \cite{andrei}.  As well as being objects of fundamental
scientific interest, 2D Wigner crystals may be of use in quantum computing
devices \cite{quantum_computing}.

At high densities the fluid ground state is paramagnetic, but at low densities
this phase is unstable to a ferromagnetic fluid.  There is some experimental
evidence that a ferromagnetic fluid is stable at densities between the
paramagnetic fluid and crystal phases \cite{ghosh}, but these data are not
conclusive.  Previous theoretical studies \cite{tanatar,rapisarda} have
reported a region of stability for the ferromagnetic fluid.  However, the
energy differences between the various phases are small, and highly accurate
calculations are required to resolve them.  Another issue of recent interest
is the possible existence of intermediate phases in the vicinity of the
fluid--crystal transition \cite{spivak,hexatic,supersolid,falakshahi}.  In
this paper we address the issues of the stability of the ferromagnetic fluid,
the fluid--crystal transition, the magnetic behavior of the crystal, and the
possibility of a ``hybrid'' phase of the type proposed in Ref.\
\cite{falakshahi}.

We have performed quantum Monte Carlo (QMC) \cite{foulkes_2001,ceperley_1980}
calculations for the 2D fluid and Wigner crystal phases, achieving much
smaller statistical error bars than previous studies
\cite{tanatar,rapisarda,attaccalite}.  We used the variational and diffusion
quantum Monte Carlo (VMC and DMC) methods as implemented in the
\textsc{casino} code \cite{casino}.  DMC is the most accurate method available
for studying quantum many-body systems such as electron gases. Fermionic
symmetry is imposed via the fixed-node approximation \cite{anderson_1976}, in
which the nodal surface is constrained to equal that of a trial wave function.
The Ewald method was used to calculate the electron-electron interaction
energy \cite{wood}.  We used extremely flexible Slater-Jastrow-backflow
\cite{backflow} wave functions, and we optimized the crystal orbitals by
directly minimizing the DMC energy.  Our fixed-node DMC energies are therefore
more accurate than those of earlier calculations.  Finally, we have dealt with
finite-size effects in a more sophisticated fashion than previous studies
\cite{ndd_fs}.

It is well-established that the triangular crystal has a considerably lower
energy than competing lattices in the density range of interest.  A triangular
lattice was therefore used in all our crystal calculations.
Antiferromagnetism is frustrated on a triangular lattice, and
antiferromagnetic crystals are expected to form a spin liquid
\cite{bernu}. Instead, we have considered antiferromagnetic crystals
consisting of alternating lines of spin-up and spin-down electrons.

We used the Jastrow factor proposed in Ref.\ \cite{ndd_jastrow}.  Plane-wave
orbitals $\exp(i{\bf G}\cdot {\bf r})$ were used in the Slater wave function
in our fluid calculations, while identical Gaussian orbitals $\exp(-Cr^2)$
centered on lattice sites were used in our crystal calculations.  The Slater
wave function was evaluated at quasiparticle coordinates defined by the
backflow (BF) transformation described in Ref.\ \cite{backflow}.  The free
parameters in the Jastrow factor and BF function were optimized within VMC by
minimizing the energy \cite{umrigar_emin}.  The BF functions for antiparallel
spins are much larger than those for parallel spins, which are already kept
apart by Pauli exclusion.  BF is less important in the crystal because the
localization of the orbitals keeps the electrons apart.  For example, at
$r_s=30$ a.u.\ \cite{note1}, BF lowers the DMC energy of the 42-electron
paramagnetic fluid by 0.000036(3) a.u.\ per electron, whereas it lowers the
energy of the 64-electron ferromagnetic Wigner crystal by only 0.0000023(3)
a.u.\ per electron.  Time-step bias was removed from our final DMC energies by
linear extrapolation to zero time step \cite{EPAPS}.  A target population of
at least 1500 configurations was used in our production runs, making
population-control bias negligible \cite{EPAPS}.

We optimized the Gaussian exponent $C$ in the crystal orbitals by minimizing
the Slater-Jastrow DMC energy.  At $r_s=40$ a.u., the DMC-optimized exponent
of the ferromagnetic crystal ($C_{\rm DMC}^{\rm F} \approx 0.0003$ a.u.)\ is
smaller than the exponent obtained by minimizing the VMC energy ($C_{\rm
VMC}^{\rm F} \approx 0.0006$ a.u.\ with our Jastrow factor), which is, in
turn, substantially smaller than the exponent within either Hartree or
Hartree-Fock theory ($C_{\rm H}^{\rm F}=0.0019$ a.u.)\ \cite{trail_wigner}.
The DMC energy depends much less sensitively than the VMC energy on the value
of $C$ \cite{EPAPS}.  However, the energy difference between DMC results
obtained with $C=C_{\rm DMC}^{\rm F}$ and $C=C_{\rm VMC}^{\rm F}$ is
significant, while the fixed-node error that results from using $C=C_{\rm
H}^{\rm F}$ is very substantial.  We find that the DMC-optimized exponents of
64-electron crystals at different densities are given by $C_{\rm DMC}^{\rm F}
= 0.071 \, r_s^{-3/2}$ and that the DMC-optimized exponents for 16-, 64-, and
196-electron crystals at $r_s=40$ a.u.\ are very similar.  We have therefore
used $C=C_{\rm DMC}^{\rm F}$ at all densities and system sizes. For
antiferromagnetic crystals we have used $C=C_{\rm DMC}^{\rm AF} = 0.082 \,
r_s^{-3/2}$.

Simulations were performed with up to 162, 109, 100, and 121 electrons for the
paramagnetic fluid, fully spin-polarized (ferromagnetic) fluid,
antiferromagnetic crystal, and ferromagnetic crystal, respectively
\cite{EPAPS}.  We eliminated single-particle finite-size effects from the
fluid energies by twist averaging within the canonical ensemble
\cite{lin_twist_av}.  Every so often during VMC or DMC simulations, an offset
to the grid of ${\bf G}$ vectors was chosen at random in the first Brillouin
zone of the simulation cell, the lowest-energy plane-wave orbitals were
selected, and a short period of re-equilibration was carried out.  The
finite-size errors in the energy per electron resulting from the compression
of the exchange-correlation hole into the simulation cell and the neglect of
long-ranged correlations in the kinetic energy fall off as $N^{-5/4}$
\cite{ndd_fs}.  We therefore extrapolated our fluid energies to infinite
system size by fitting our data at each density to $E_N = E_\infty-c/N^{5/4}$,
where $E_N$ is the energy per electron of the $N$-electron system and $c$ and
$E_\infty$ are fitting parameters \cite{EPAPS}. This differs from the form of
bias that has been incorrectly assumed in previous studies of the 2D HEG
\cite{rapisarda,tanatar,attaccalite}.

Our DMC results for the different phases are given in Table
\ref{table:DMC_extrap_results}.  Our Wigner-crystal energy data were fitted to
the first five terms in the low-density expansion of the crystal energy
($E=c_1 r_s^{-1}+c_{3/2} r_s^{-3/2}+c_2 r_s^{-2}+c_{5/2} r_s^{-5/2}+c_3
r_s^{-3}$) \cite{ceperley_1978}.  The first term is the Madelung energy of the
static lattice, while the second is the quasiharmonic zero-point phonon
energy.  The corresponding coefficients can be determined analytically:
$c_1=-1.106103$ and $c_{3/2}=0.814$ \cite{bonsall}.  The remaining three
coefficients were determined by fitting to our QMC data, giving
$c_2=0.113743$, $c_{5/2}=-1.184994$, and $c_3=3.097610$ for the ferromagnetic
crystal and $c_2=0.2662977$, $c_{5/2}=-2.63286$, and $c_3=6.246358$ for the
antiferromagnetic crystal.  We fitted our fluid energy data to the
parameterization of the correlation energy suggested by Rapisarda and Senatore
[Eq.\ (28) of Ref.\ \onlinecite{rapisarda}].  For the paramagnetic fluid we
find $a_0=-0.1863052$\ a.u., $a_1=6.821839$, $a_2=0.155226$, and
$a_3=3.423013$, where $a_0$, $a_1$, $a_2$, and $a_3$ are the free parameters
in Rapisarda and Senatore's fitting form. For our ferromagnetic data we obtain
$a_0=-0.2909102$\ a.u., $a_1=-0.6243836$, $a_2=1.656628$, and $a_3=3.791685$.
We find a greater difference between the energies of the paramagnetic and
ferromagnetic Fermi fluids at $r_s=20$ a.u.\ than Attaccalite \textit{et al.}
\cite{attaccalite}, resulting in somewhat smaller values of the spin
susceptibility.

\begin{table}
\begin{center}
\begin{tabular}{ccccc}
\hline \hline

& \multicolumn{4}{c}{DMC energy ($10^{-2}$ a.u.\ / elec.)} \\

\raisebox{1ex}[0pt]{$r_s$} & Ferro.\ crystal & Antif.\ crys. & Para.\ fluid &
Ferro.\ fluid \\

\hline

15 & $-5.966\,5(1)$ & $\cdots$ & $\cdots$ & $\cdots$ \\

20 & $-4.619\,5(2)$ & $-4.622\,9(2)$ & $-4.630\,5(4)$ & $-4.621\,3(3)$ \\

25 & $-3.773\,1(2)$ & $-3.775\,1(3)$ & $-3.777\,4(2)$ & $-3.774\,0(2)$ \\

30 & $-3.191\,7(2)$ & $-3.192\,2(2)$ & $-3.192\,6(1)$ & $-3.191\,3(1)$ \\

35 & $-2.766\,9(1)$ & $-2.767\,2(1)$ & $-2.766\,5(1)$ & $-2.765\,7(1)$ \\

40 & $-2.443\,2(1)$ & $-2.443\,1(2)$ & $-2.441\,6(1)$ & $-2.441\,6(1)$ \\

45 & $-2.188\,1(1)$ & $-2.187\,5(2)$ & $\cdots$ & $\cdots$ \\

50 & $-1.981\,7(2)$ & $-1.981\,4(2)$ & $\cdots$ & $\cdots$ \\

\hline \hline
\end{tabular}
\caption{DMC energy as a function of $r_s$ for ferromagnetic and
antiferromagnetic triangular Wigner crystals, and paramagnetic and fully
ferromagnetic Fermi fluids.  All results have been extrapolated to zero time
step and infinite system size.
\label{table:DMC_extrap_results}}
\end{center}
\end{table}

\begin{figure}
\begin{center}
\includegraphics[clip,scale=0.42]{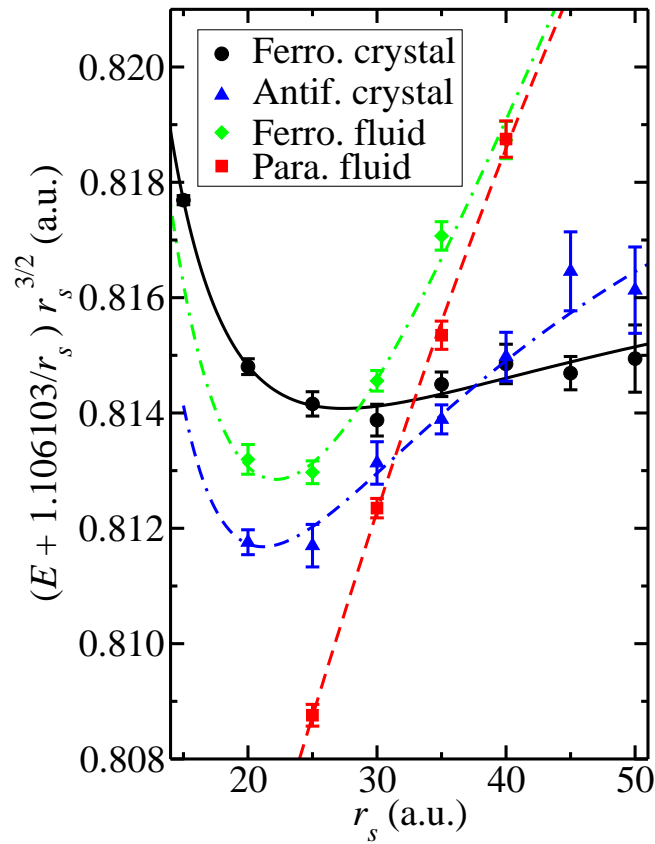}
\caption{(Color online) DMC energy as a function of $r_s$ for ferromagnetic
and antiferromagnetic triangular Wigner crystals and paramagnetic and fully
ferromagnetic Fermi fluids.  In each case the Madelung energy of a triangular
crystal has been subtracted from the energy and the result has been rescaled
by $r_s^{3/2}$ to highlight the differences between the curves.
\label{fig:HEG2D_E_v_rs}}
\end{center}
\end{figure}

The ground-state energies of the different phases of the 2D HEG are plotted
against $r_s$ in Fig.\ \ref{fig:HEG2D_E_v_rs}.  Unlike previous QMC studies,
our statistical error bars are sufficiently small that we can resolve the
energy difference between the ferromagnetic and paramagnetic fluids.  We can
also identify the crystallization density with much greater precision.
Previous studies found crystallization to occur at $r_s=37(5)$ a.u.\
\cite{tanatar} and $r_s=34(4)$ a.u.\ \cite{rapisarda}.  We find the
crystallization density to be $r_s=31(1)$\ a.u., but the transition is from
the paramagnetic fluid to an antiferromagnetic crystal, not the ferromagnetic
fluid to the ferromagnetic crystal as found in the previous studies.  Our
calculations locate the density at which the paramagnetic fluid becomes
unstable to the ferromagnetic fluid at about $r_s=40$ a.u., but at this
density the Wigner crystal is more stable than either fluid phase, so we do
not find a region of stability for the ferromagnetic fluid.  We have looked
for a region of stability for a partially spin-polarized fluid by calculating
the energy of a fluid of spin polarization $\zeta=2/5$ at $r_s=35$ a.u.  The
DMC energy, extrapolated to zero time step and infinite size, is
$-0.027666(1)$ a.u.\ per electron, which is not significantly less than the
energies of the paramagnetic and ferromagnetic fluids, but is significantly
higher than the crystal energy (see Table \ref{table:DMC_extrap_results}).  It
is therefore unlikely that a region of stability exists for a partially
spin-polarized fluid. We do, however, find a transition from an
antiferromagnetic Wigner crystal to a ferromagnetic one at $r_s=38(5)$ a.u.
This is considerably higher than the density at which Bernu \textit{et al.}\
\cite{bernu} found a transition from a spin liquid to a ferromagnetic crystal
using a multispin exchange model ($r_s=175$ a.u.). An experimental result for
the crystallization density is $r_s = 35.1(9)$ a.u.\ \cite{yoon}, which is
somewhat lower than the QMC crystallization density.  This suggests that the
ideal 2D HEG is not a perfect model for electron layers in real semiconductor
devices.

It has been argued that the transition from a 2D Fermi fluid to a Wigner
crystal cannot be first order, because, at the transition density, it is
energetically favorable to create boundaries between macroscopically separated
regions of fluid and crystal \cite{spivak}.  The energy of the Bose fluid is
substantially lower than that of the Fermi fluid in the vicinity of the
crystallization density in 2D, unlike 3D \cite{depalo}.  There is therefore
more scope for interesting phase behavior in 2D\@.  Various intermediate
phases have been proposed in the literature, such as a hexatic phase with
orientational but not translational order \cite{hexatic}, a supersolid phase
\cite{supersolid}, or a microemulsion phase \cite{spivak}.  Falakshahi and
Waintal \cite{falakshahi} have suggested that a ``hybrid'' phase is stable in
the vicinity of the transition density from a ferromagnetic fluid to a
ferromagnetic Wigner crystal.  (We have found that a paramagnetic fluid is
stable at this density; however, we restrict our attention to ferromagnetic
HEGs to investigate the proposed hybrid phase.)  The hybrid phase has the same
symmetry as the Wigner crystal, but has partially delocalized orbitals.

Falakshahi and Waintal generated hybrid orbitals for the ferromagnetic
$N$-electron system by solving the Schr\"odinger equation for a single,
positively charged particle moving in a lattice of negative charges placed at
the $N$ Wigner-crystal lattice sites. If the charge of the test particle is
zero then the orbitals for the system are plane waves, i.e., fluid orbitals
are obtained.  If the charge is large then localized (crystal-like) orbitals
are obtained.  As the charge of the test particle is increased from zero,
there must come a point at which a band gap opens up between the $N$ and
$(N+1)$th states.  The set of orbitals obtained at the point at which the gap
opens correspond to the hybrid phase.  We have tried to find such a hybrid
phase by optimizing orbitals of the form
\begin{equation}
\phi({\bf r}) = A \exp(-Cr^2) + \sum_{P} c_P \sum_{{\bf G} \in P} \cos ( {\bf
G} \cdot {\bf r})
\label{eqn:hybrid_orb}
\end{equation}
centered on the crystal lattice sites, where $P$ denotes a star of
symmetry-equivalent simulation-cell ${\bf G}$ vectors.  The coefficients $\{
c_P \}$ and $C$ are optimizable parameters.  Equation (\ref{eqn:hybrid_orb})
is a general expansion of Wannier-like orbitals.  It can therefore describe
the crystal and hybrid phases, but not the fluid, since the latter corresponds
to a partially filled band.

Our QMC results for ferromagnetic HEGs at $r_s=30$ a.u.\ obtained with
different orbitals are shown in Table \ref{table:look_for_hybrid}.  Within VMC
it is possible to lower the energy significantly by optimizing the plane-wave
coefficients in Eq.\ (\ref{eqn:hybrid_orb}), apparently suggesting that
Gaussian crystal orbitals are nonoptimal.  This does not carry over to DMC,
however.  In fact it is more important to use Gaussian exponents optimized
within DMC than it is to use either plane-wave expansions in the orbitals or
BF functions, suggesting that the fixed-node errors in our crystal DMC
energies are very small.  We have searched for the hybrid phase by optimizing
wave functions with orbitals of the form given in Eq.\ (\ref{eqn:hybrid_orb})
using different starting points, but have not been able to lower the DMC
energy significantly. While this does not prove the nonexistence of the hybrid
phase, as searching for minima in a high-dimensional space is a very difficult
problem, the fact that our extensive searches have been unable to find it
makes its existence unlikely.

\begin{table}
\begin{center}
\begin{tabular}{cccr@{.}l}
\hline \hline

Method & Orbitals & BF & \multicolumn{2}{c}{Energy (a.u.\ / elec.)} \\

\hline

VMC & DMC-opt.\ Gauss.      & No  & ~~~$-0$&$031\,849\,8(1)$ \\

VMC & DMC-opt.\ Gauss.$+$PW & No  & $-0$&$031\,849\,9(1)$ \\

VMC & VMC-opt.\ Gauss.      & No  & $-0$&$031\,856\,2(1)$ \\

VMC & Gauss.$+$PW           & No  & $-0$&$031\,861\,9(3)$ \\

VMC & Gauss.$+$PW           & Yes & $-0$&$031\,871\,3(1)$ \\

DMC & Gauss.$+$PW           & Yes & $-0$&$031\,918\,0(3)$ \\

DMC & Gauss.$+$PW           & No  & $-0$&$031\,918\,4(3)$ \\

DMC & VMC-opt.\ Gauss.      & No  & $-0$&$031\,918\,4(3)$ \\

DMC & DMC-opt.\ Gauss.      & No  & $-0$&$031\,919\,0(2)$ \\

DMC & DMC-opt.\ Gauss.$+$PW & No  & $-0$&$031\,920\,1(3)$ \\

\hline \hline
\end{tabular}
\caption{Non-twist-averaged QMC energies for a ferromagnetic, 121-electron HEG
at $r_s=30$ a.u.\ obtained using different orbitals: crystal (Gaussian)
orbitals, in which the Gaussian exponent has been optimized within VMC or DMC
and ``hybrid'' (Gauss.$+$plane-wave) orbitals of the form given in Eq.\
(\ref{eqn:hybrid_orb}) with the coefficients of 20 stars of ${\bf G}$ vectors
optimized within VMC\@.  The orbitals were centered on the lattice sites of a
triangular crystal. VMC-optimized BF functions were used in some of the
calculations.  DMC energies were obtained using a time step of 1 a.u.  The
non-twist-averaged DMC energy of the ferromagnetic Fermi fluid at this system
size and density [$-0.0319354(5)$ a.u.\ per electron] is relatively low due to
single-particle finite-size effects, so that, if anything, finite-size effects
are expected to favor the hybrid phase.
\label{table:look_for_hybrid}}
\end{center}
\end{table}

In summary, we have studied the zero-temperature phase behavior of the 2D HEG
using QMC\@.  We find a transition from a paramagnetic fluid to an
antiferromagnetic triangular Wigner crystal at $r_s=31(1)$\ a.u.  We find no
region of stability for a ferromagnetic fluid, although we find that the
paramagnetic fluid is unstable to the ferromagnetic fluid at about $r_s=40$\
a.u.\ and the antiferromagnetic crystal is unstable to the ferromagnetic
crystal at $r_s=38(5)$ a.u.  We find no evidence for the existence of hybrid
phases of the type suggested by Falakshahi and Waintal \cite{falakshahi}, but
we cannot, of course, rule out the existence of other types of intermediate
phase.

\begin{acknowledgments}
We acknowledge financial support from Jesus College, Cambridge and the UK
Engineering and Physical Sciences Research Council (EPSRC)\@.  Computing
resources were provided by the Cambridge High Performance Computing Service
and HPCx.
\end{acknowledgments}


\begin{thebibliography}{99}
\bibitem{note1} Densities are given in terms of the radius $r_s$ of the circle
that contains one electron on average.  We use Hartree atomic units
($\hbar=|e|=m_e=4\pi\epsilon_0=1$) throughout.
\end{thebibliography}
\end{document}